\documentclass[10pt]{article}
\usepackage{geometry}
\usepackage{graphicx}
\usepackage{amsmath,amssymb}
\usepackage{cite}
\newcommand{\lw}[1]{\smash{\lower 1.5ex\hbox{#1}}}
\newcommand{\ri}[1]{\smash{\raise 1.5ex\hbox{#1}}}
\newcommand{\riw}[1]{\smash{\raise 3.0ex\hbox{#1}}}
\newcommand{\diff}[2]{\frac{\partial #1}{\partial #2}}
\newcommand{\mapright}[1]{\smash{\mathop{\hbox to 3.0cm{\rightarrowfill}}\limits^{\displaystyle #1}}}

\begin{document}

\title{Analysis of Bose-Einstein correlation at 7 TeV by\\ LHCb collaboration based on stochastic approach}

\author{Takuya Mizoguchi$^{1}$ and Minoru Biyajima$^{2}$\\
{\small $^{1}$National Institute of Technology, Toba College, Toba 517-8501, Japan}\\
{\small $^{2}$Department of Physics, Shinshu University, Matsumoto 390-8621, Japan}}

\maketitle

\begin{abstract}
The Bose-Einstein correlation (BEC) in forward region ($2.0<\eta<4.8$) measured at 7 TeV in the Large Hadron Collider (LHC) by the LHCb collaboration is analyzed using two conventional formulas of different types named CF$_{\rm I}$ and CF$_{\rm II}$. The first formula is well known and contains the degree of coherence ($\lambda$) and the exchange function $E_{\rm BE}^2$ from the BE statistics. The second formula is an extended formula (CF$_{\rm II}$) that contains the second degree of coherence $\lambda_2$ and the second exchange function $E_{\rm BE_2}^2$ in addition to CF$_{\rm I}$. To examine the physical meaning of the parameters estimated by CF$_{\rm II}$, we analyze the LHCb BEC data by using a stochastic approach of the three-negative binomial distribution and the three-generalized Glauber-Lachs formula. Our results reveal that the BEC at 7 TeV consisted of three activity intervals defined by the multiplicity $n$ ([8, 18], [19, 35], and [36, 96]) can be well explained by CF$_{\rm II}$.
\end{abstract}


\section{\label{sec1}Introduction}
An analysis of the Bose-Einstein correlation (BEC) reported by LHCb collaboration~\cite{Aaij:2017oqu} is an interesting prospect because the data are measured in the forward region ($2.0<\eta <4.8$) at 7 TeV, and the number of data points is large (390). In the report of the BEC, the authors used the following conventional formula:
\begin{eqnarray}
{\rm CF_I} = 1.0 + \lambda_1 E_{\rm BE}^2,
\label{eq1}
\end{eqnarray}
where $\lambda_1$ and $E_{\rm BE}^2$ are the degree of coherence and the exchange function between the same charged pions according to the BE statistics, respectively. The exchange function can be expressed as follows:
\begin{eqnarray}
E_{\rm BE}^2 =\ \left\{
\begin{array}{l}
\exp(-(RQ)) \mbox{ (Exponential function) (E)},\medskip\\
\exp(-(RQ)^2) \mbox{ (Gaussian distribution) (G)},
\end{array}
\right.
\label{eq2}
\end{eqnarray}
where $Q=\sqrt{-(p_1-p_2)^2}$ is the root of the momentum transfer square. 

For applications of Eqs.~(\ref{eq1}) and (\ref{eq2}), we introduce a normalization factor and a long-range correlation effect of $c(1+\delta)$. The results generated using Eqs.~(\ref{eq1}) and (\ref{eq2}) with the exponential function (E) and the Gaussian distribution (G) are presented in Table~\ref{tab1}. As can be seen in the table, the $\chi^2$ values are greater than the data points. In this study, we investigate whether CF$_{\rm I}$ (Eq.~(\ref{eq1})) could describe the BEC measured by the LHCb collaboration. This means that we also have to examine the extended formulas presented in~\cite{Mizoguchi:2010vc,Biyajima:2018abe,Mizoguchi:2019cra,Biyajima:2019wcb}.\footnote{In Ref.~\cite{Biyajima:2018abe}, for $N^{\rm BG}$, an identical separation between two ensembles with $\alpha_1$and $\alpha_2$ is assumed. For no-separation between them, the following formula is obtained:
$$
N^{\rm (2+:\,2-)}/N^{\rm BG} = 1 + (a_1/s)E_1^2 +(a_2/s)E_2^2,
$$
where $s=a_1+a_2 =\alpha_1\langle n_1\rangle^2 + \alpha_2\langle n_2\rangle^2$ (see succeeding Ref.~\cite{Mizoguchi:2019cra}) works in the present analysis.
}

\begin{table}[h]
\centering
\caption{\label{tab1}Results obtained using Eqs.~(\ref{eq1}) and (\ref{eq2}).} 
\vspace{1mm}
\begin{tabular}{cccccc}
\hline
Activity interval & Eq.~(\ref{eq2}) & $R$ (fm) & $\lambda$ & $\delta\,(\times 10^{-2}\,{\rm GeV^{-1}})$ & $\chi^2$/ndf\\
\hline
Low
& (E)
& $1.01\pm 0.01$
& $0.72\pm 0.01$
& $8.9\pm 0.2$
& $591/386$\\
$[8,\,18]$
& (G)
& $0.68\pm 0.01$
& $0.36\pm 0.00$
& $5.4\pm 0.2$
& $1979/386$\\
\hline
Medium
& (E)
& $1.48\pm 0.02$
& $0.63\pm 0.01$
& $4.9\pm 0.1$
& $623/386$\\
$[19,\,35]$
& (G)
& $0.88\pm 0.01$
& $0.31\pm 0.00$
& $3.7\pm 0.1$
& $1785/386$\\
\hline
High
& (E)
& $1.81\pm 0.03$
& $0.57\pm 0.01$
& $2.6\pm 0.1$
& $621/386$\\
$[36,\,96]$
& (G)
& $1.02\pm 0.01$
& $0.27\pm 0.00$
& $2.0\pm 0.1$
& $1348/386$\\
\hline
\end{tabular}
\end{table}

To discover how the estimated parameters ($\lambda$ and $R$) are different from those in Eq.~(\ref{eq1}), we apply the following extended conventional formula (CF$_{\rm II}$) for the analysis of the BEC:
\begin{eqnarray}
{\rm CF_{II}} = 1.0 + \lambda_1 E_{\rm BE_1}^2 + \lambda_2 E_{\rm BE_2}^2,
\label{eq3}
\end{eqnarray}
In the above formula, $\lambda_2$ is the second degree of coherence. The second exchange function is highly similar to Eq.~(\ref{eq2}) but gives a different value for the interaction range $R_2$. In applying Eq.~(\ref{eq3}) to the BEC, we select the geometrical combinations [E + E], [G + G], [G + E], and [E + G]. We then have to perform 1,000 trials using the CERN MINUIT program, because CF$_{\rm II}$ contains additional parameters ($R_2$ and $\lambda_2$). Our use of CF$_{\rm II}$ is described in more detail in appendix~\ref{secA}. According to our CF$_{\rm II}$ calculations, we obtain several ensembles specified by $\chi^2$ values and degree of emergence (d.e.) which is defined by the ratio of the number of the $\chi^2$ ensembles with the same $\chi^2$ value to the 1,000 trials.
$$
\mbox{(d.e.)} = \frac{\mbox{the number of ensembles with the same }\chi^2}{\mbox{the number of trials (\mbox{1,000})}}
$$

\begin{figure}[htbp]
  \centering
  \includegraphics[width=0.48\columnwidth]{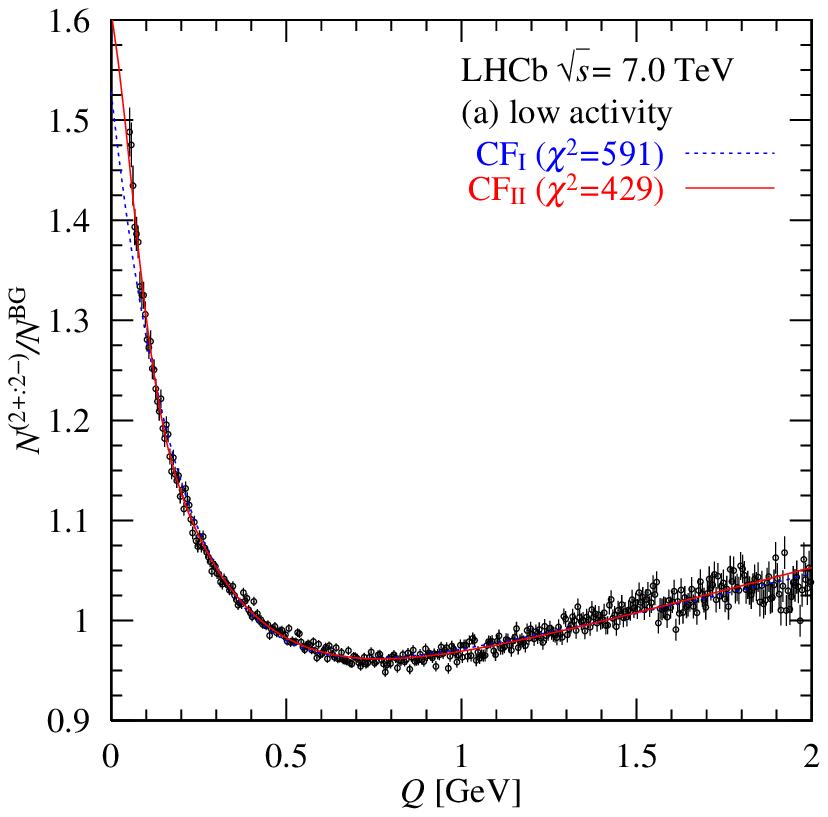}
  \includegraphics[width=0.48\columnwidth]{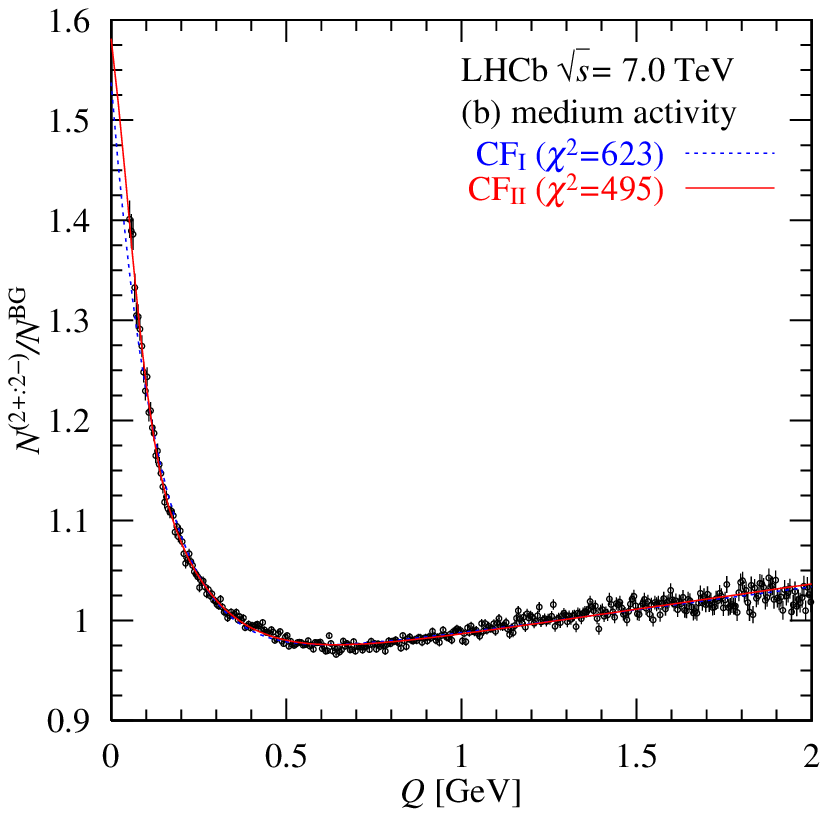}\\
  \includegraphics[width=0.48\columnwidth]{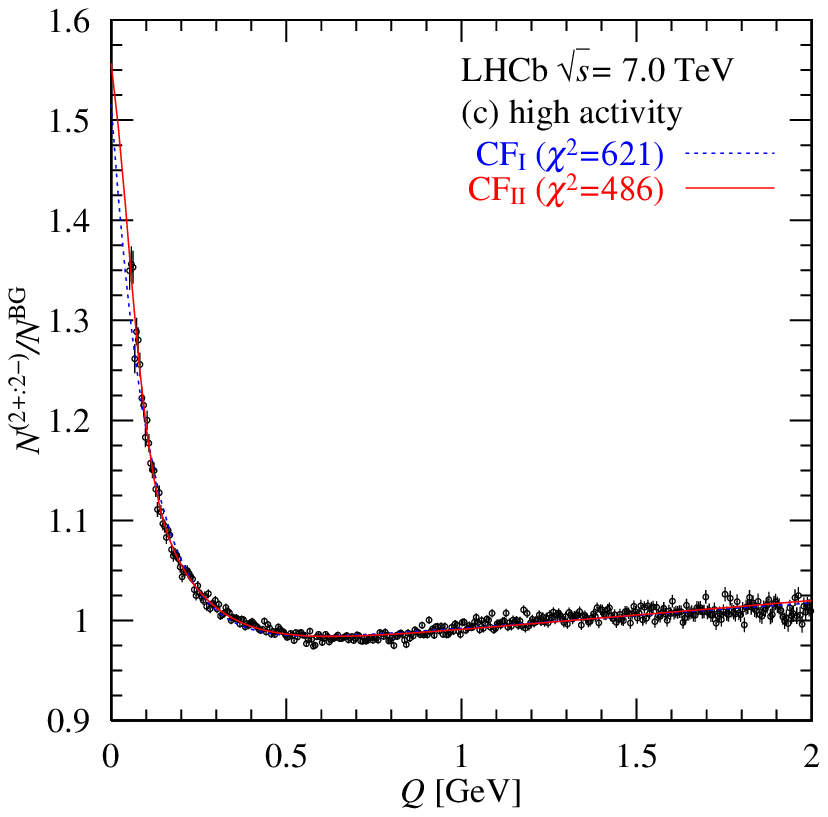}
  \caption{\label{fig1}Analysis of the BEC at three activity intervals using Eq.~(\ref{eq1}) with the exponential function (E) and Eq.~(\ref{eq3}) with [E + G]. Values of $\chi^2$ for the [E + G] configuration are the smallest ones among the four configurations ([E + E], [E + G], [G + E], and [G + G]).}
\end{figure}

\begin{figure}[htbp]
  \centering
  \includegraphics[width=0.48\columnwidth]{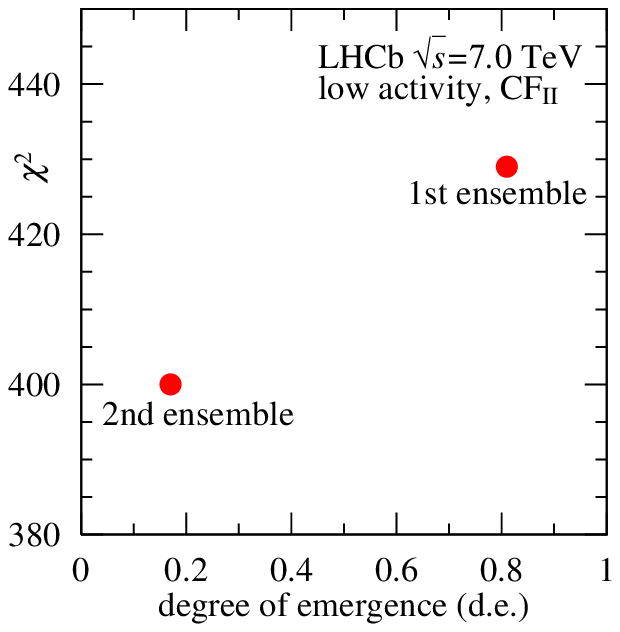}
  \caption{\label{fig2}Ensembles specified by ((d.e.), $\chi^2$).}
\end{figure}

As is seen in Fig.~\ref{fig2}, we observe two ensembles specified by ((d.e.), $\chi^2$ values). Thus we compare $R$'s by CF$_{\rm I}$ in Table~\ref{tab1} and $R_1$'s by CF$_{\rm II}$ in Table~\ref{tab2}, because the second term of the right hand side (RHS) of CF$_{\rm II}$ may correspond to the second one in CF$_{\rm I}$. In other words, the third term of the RHS of CF$_{\rm II}$ is regarded as the correction term to the second term, because $\lambda_1$'s $> \lambda_2$'s. Our result by Eq.~(\ref{eq3}) is given in Fig.~\ref{fig1} and Table~\ref{tab2}.

\begin{table}[h]
\centering
\caption{\label{tab2}Results obtained using Eqs.~(\ref{eq2}) and (\ref{eq3}).} 
\vspace{1mm}
\begin{tabular}{ccccccc}
\hline
$\!\!\!$Activity interval$\!\!\!\!\!\!$ & $R_1$ (fm) & $\lambda_1$ & $R_2$ (fm) & $\lambda_2$ & $\delta\,(\times 10^{-2})$ & $\chi^2$/ndf (d.e.)\\
\hline
Low
& $0.85\pm 0.02$ (E)
& $0.63\pm 0.01$
& $2.08\pm 0.15$ (G)
& $0.22\pm 0.02$
& $10.5\pm 0.3$
& $429/384$ (0.81)\\
\hline
Medium
& $1.26\pm 0.03$ (E)
& $0.50\pm 0.02$
& $2.27\pm 0.13$ (G)
& $0.19\pm 0.02$
& $5.4\pm 0.1$
& $495/384$ (0.77)\\
\hline
High
& $1.43\pm 0.04$ (E)
& $0.38\pm 0.02$ 
& $2.44\pm 0.12$ (G)
& $0.24\pm 0.02$
& $3.0\pm 0.1$
& $486/384$ (0.80)\\
\hline\hline

\multicolumn{7}{l}{From top to bottom: i) low, ii) medium, and iii) high activity}\\
$\!\!$i) CF$_{\rm II}$
& $2.07\pm 0.10$ (E)
& $0.81\pm 0.02$
& $0.42\pm 0.01$ (G)
& $0.18\pm 0.01$
& $9.4\pm 0.3$
& $400/384$ (0.17)\\

$\!\!$ii) CF$_{\rm II}$
& $2.74\pm 0.12$ (E)
& $0.80\pm 0.03$
& $0.54\pm 0.01$ (G)
& $0.12\pm 0.01$
& $5.2\pm 0.1$
& $455/384$ (0.23)\\

$\!\!$iii) CF$_{\rm II}$
& $\!\!\!\!\left\{\!\!\begin{array}{l}
3.42\pm 0.17\ ({\rm E})\!\!\\
1.81\pm 0.03\ ({\rm E})\!\!
\end{array}\right.\!\!$
& $\!\!\begin{array}{l}
0.85\pm 0.04\!\!\\
0.57\pm 0.01\!\!
\end{array}$
& $\!\!\begin{array}{l}
0.60\pm 0.02\ ({\rm G})\!\!\\
1.3\pm 31.2\ ({\rm G})\!\!
\end{array}$
& $\!\!\begin{array}{l}
0.09\pm 0.01\!\!\\
0.00\pm 0.00\!\!
\end{array}$
& $\!\!\begin{array}{l}
2.9\pm 0.1\!\!\\
2.6\pm 0.1\!\!
\end{array}$
& $\!\!\begin{array}{l}
461/384\ (0.16)\!\!\\
621/384\ (0.04)\!\!
\end{array}$\\

\hline
\end{tabular}
\end{table}


As the next step, we have to elucidate physical meaning of $\lambda_1$ and $\lambda_2$ in a different point of view. For our purpose, we would like to consider the stochastic approach, first of all, the three-negative binomial distribution (T-NBD). Actually, in Ref.~\cite{Biyajima:2018abe,Mizoguchi:2019cra,Biyajima:2019wcb}, we have shown that the degrees of coherence ($\lambda_1$ and $\lambda_2$) are calculated by stochastic approach, i.e., the three-negative binomial distribution (T-NBD). The T-NBD formula is expressed as
\begin{eqnarray}
P(n,\, \langle n\rangle) &=& \sum_{i=1}^3 \alpha_i P_{{\rm NBD}_i}(n,\,\langle n_i\rangle,\,k_i),\nonumber\\
P_{\rm NBD}(n,\,\langle n\rangle,\,k) &=& \frac{\Gamma (n+k)}{\Gamma (n+1)\Gamma (k)}\frac{(\langle n\rangle/k)^n}{(1+\langle n\rangle/k)^{n+k}}
\label{eq4}
\end{eqnarray}
where $\alpha_1+\alpha_2+\alpha_3=1.0$. $\langle n\rangle$ and $k$ are the average multiplicity and the intrinsic parameters, respectively.

In this paper, we use Eq.~(\ref{eq4}) to analyze the multiplicity distribution (MD) data obtained by the LHCb collaboration~\cite{Aaij:2011yj,Aaij:2014pza}. The moments of Eq.~(\ref{eq4}) can be calculated as follows:
\begin{eqnarray}
\langle n_i\rangle &=& \sum_{n=0}^{\infty} P_{{\rm NBD}_i}(n,\,\langle n_i\rangle,\,k_i)n,\nonumber\\
\langle n_i(n_i-1)\rangle &=& \sum_{n=0}^{\infty} P_{{\rm NBD}_i}(n,\,\langle n_i\rangle,\,k_i)n(n-1).
\label{eq5}
\end{eqnarray}
The BEC is calculated using Eq.~(\ref{eq5}) and the weight factor $\alpha_i$'s.

Before performing concrete calculations using T-NBD, we have to demonstrate physical correspondence between the three components and a classification of the LHC collisions. Analyses by T-NBD performed in~\cite{Zborovsky:2013tla}, \cite{Biyajima:2018abe}, and \cite{Biyajima:2019wcb} have revealed that a variety of MDs at the LHC can be explained with very small $\chi^2$ values. This may be attributed to the fact that the first component with $\alpha_1$ corresponds to non-diffractive dissociation (ND), the second one with $\alpha_2$/or $\alpha_3$ does to single diffractive dissociation (SD), and the third one with $\alpha_3$/or $\alpha_2$ does to double diffractive dissociation (DD)~\cite{GrosseOetringhaus:2009kz,Navin:2010kk,ATLAS:2010mza}:
\begin{eqnarray*}
\left\{\begin{array}{l}
\mbox{The first component with $\alpha_1$} \leftrightarrow \mbox{ Non-diffractive dissociation (ND),}\\
\mbox{The second one with $\alpha_2$/or $\alpha_3$} \leftrightarrow \mbox{ Single diffractive dissociation (SD),}\\
\mbox{The third one with $\alpha_3$/or $\alpha_2$} \leftrightarrow \mbox{ Double diffractive dissociation (DD).}
\end{array}\right.
\end{eqnarray*}

Moreover, T-NBD demonstrates an interesting oscillatory pattern that can be explained by the stochastic theory~\cite{Zborovsky:2018vyh,Wilka:2016ufh,Wilk:2018fhw}. Those previous findings support the hypothesis that T-NBD would work well in an analysis of MD from the LHC.
\\

The second section presents our analysis of the MD measured at 7 TeV using T-NBD. Using T-NBD parameters, we are able to determine the degrees of coherence necessary for the analysis of the BEC. The third section presents our analysis of the BEC using the formulas determined in the second section, and the fourth section presents concluding remarks and discussions. Appendix~\ref{secA} presents the details on our use of CF$_{\rm II}$. Appendix~\ref{secB} presents our calculations of the MDs at three activity intervals using T-NBD. Appendix~\ref{secC} presents the framework of the three-generalized Glauber-Lachs (T-GGL) formula~\cite{Biyajima:1982un,Biyajima:1983qu,Biyajima:1990ku,Shimoda:1957aa} along with its use in the calculation of several physical quantities. Appendix~\ref{secD} presents the optics branching equation~\cite{Shimoda:1957aa} and the quantum chromodynamics (QCD) equation~\cite{Biyajima:1984aq,Biyajima:1984ay}.


\section{\label{sec2}Analysis of MD using T-NBD}
\begin{figure}[htbp]
  \centering
  \includegraphics[width=0.48\columnwidth]{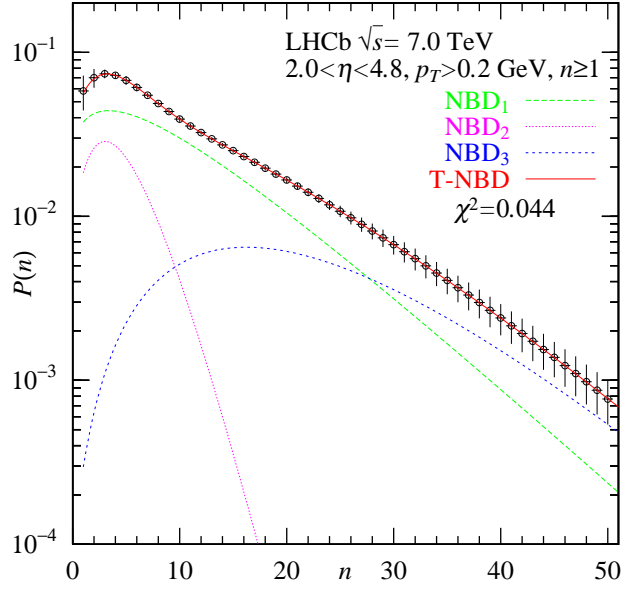}
  \caption{\label{fig3}Analysis of the MD using Eq.~(\ref{eq4}).}
\end{figure}

The multiplicity distribution (MD) measured by the LHCb collaboration is reported in~\cite{Aaij:2011yj} and \cite{Aaij:2014pza}. Making use of Eq.~(\ref{eq4}), we are able to analyze these data. The results obtained using the minimum $\chi^2$ value (0.044) are shown in Fig.~\ref{fig3}. 

\begin{table}[htbp]
\centering
\caption{\label{tab3}Estimated MD parameters using T-NBD (Eq.~(\ref{eq4})). $\chi^2=0.044$.} 
\vspace{1mm}
\begin{tabular}{cccc}
\hline
$i$ & $\alpha_i$ & $\langle n_i\rangle$ & $k_i$\\
\hline
1
& $0.648\pm 0.102$
& $10.32\pm 1.63$ 
& $1.60\pm 0.44$\\
2
& $0.180\pm 0.059$
& $4.42\pm 0.53$  
& $5.06\pm 3.78$\\
3
& $0.173\pm 0.059$
& $22.1\pm 3.6$ 
& $4.10\pm 1.61$\\
\hline
\end{tabular}
\end{table}

Making use of the values presented in Table~\ref{tab3}, we are able to calculate several moments throughout the three activity intervals. These are presented in Appendix \ref{secB}.

Because the second moment in the low activity interval is expressed as the first T-NBD component ($i=1$) (with $a=+$ or $a=-$), the following holds: 
\begin{eqnarray}
\alpha_1\langle n_1^a(n_1^a-1)\rangle_{\rm low} &=& \alpha_1\frac{\langle n_1^a(n_1^a-1)\rangle_{\rm low}}{\langle n_1(n_1-1)\rangle_{\rm tot}}\langle n_1(n_1-1)\rangle_{\rm tot}\nonumber\\
&=& \alpha_1\beta_1 \frac{\langle n_1\rangle^2}4 \left(1+\frac 2{k_1}\right),\nonumber\\
&=& \tilde{\alpha}_1\left(1+\frac 2{k_1}\right),
\label{eq6}
\end{eqnarray}
where the ratio $\beta_1= \langle n_1^a(n_1^a-1)\rangle_{\rm low}/\langle n_1(n_1-1)\rangle_{\rm tot}$ is playing a role of the weight factor. We can also calculate similar quantities for the second and third components ($i=2,\,3$). These calculations must be introduced because it is not possible to discover the relations between $\langle n_1^a(n_1^a-1)\rangle$ and $k_1$. In our analysis of the MD for the same charged particles, we perform the following replacements (see~\cite{Biyajima:2018abe,Mizoguchi:2019cra,Biyajima:2019wcb}):
$$
\langle n_i\rangle \to \langle n_i\rangle/2\ {\rm and}\ 1/k_i \to 2/k_i\ (i=1,\,2,\,3).
$$
Moreover, to describe the $Q$-dependence of the BEC, we have to use the exchange functions $E_{\rm BE_1}^2$, $E_{\rm BE_2}^2$, and $E_{\rm BE_3}^2$. We thus obtain the following formula with $s=\tilde{\alpha}_1+\tilde{\alpha}_2+\tilde{\alpha}_3$:
\begin{eqnarray}
{\rm BEC}_{\rm (T\mathchar`-N)} &=& 1.0 + \frac 1s \sum_{i=1}^3 \tilde{\alpha}_i\frac 2{k_i} E_{{\rm BE}_i}^2\nonumber\\
&=& 1.0 + \sum_{i=1}^3 \lambda_i^{\rm (T\mathchar`-N)} E_{{\rm BE}_i}^2
\label{eq7}
\end{eqnarray}
where $\lambda_i^{\rm (T\mathchar`-N)}= (\tilde{\alpha}_i/s)\cdot 2/k_i$ $(i=1,\,2,\,3)$.

\begin{table}[h]
\centering
\caption{\label{tab4}Three degrees of coherence in the three activity intervals.}
\vspace{1mm}
\begin{tabular}{cccc}
\hline
\vspace{-3mm}\\
Activity interval & $\lambda_1^{\rm (T\mathchar`-N)}$ & $\lambda_2^{\rm (T\mathchar`-N)}$ & $\lambda_3^{\rm (T\mathchar`-N)}$\\
\hline
Low
& 0.831
& \multicolumn{2}{c}{$\underbrace{0.022\quad 0.135}_{\mbox{0.157}}$}\\
Medium
& 0.557
& 0
& 0.270\\
High
& 0.333
& 0
& 0.358\\
\hline
\end{tabular}
\end{table}

Making use of the results for the parameters ($\lambda_i^{\rm (T\mathchar`-N)}$'s) presented Table~\ref{tab4}, we are able to analyze the BEC. The results are presented in Fig.~\ref{fig4} and Table~\ref{tab5}. At the low activity interval, the contribution of $\lambda_2^{\rm (T\mathchar`-N)}$ is absorbed by $\lambda_3^{\rm (T\mathchar`-N)}$ because it is very small (0.022).


\section{\label{sec3}Analysis of BEC by Eq.~(\ref{eq7})}
\begin{figure}[htbp]
  \centering
  \includegraphics[width=0.48\columnwidth]{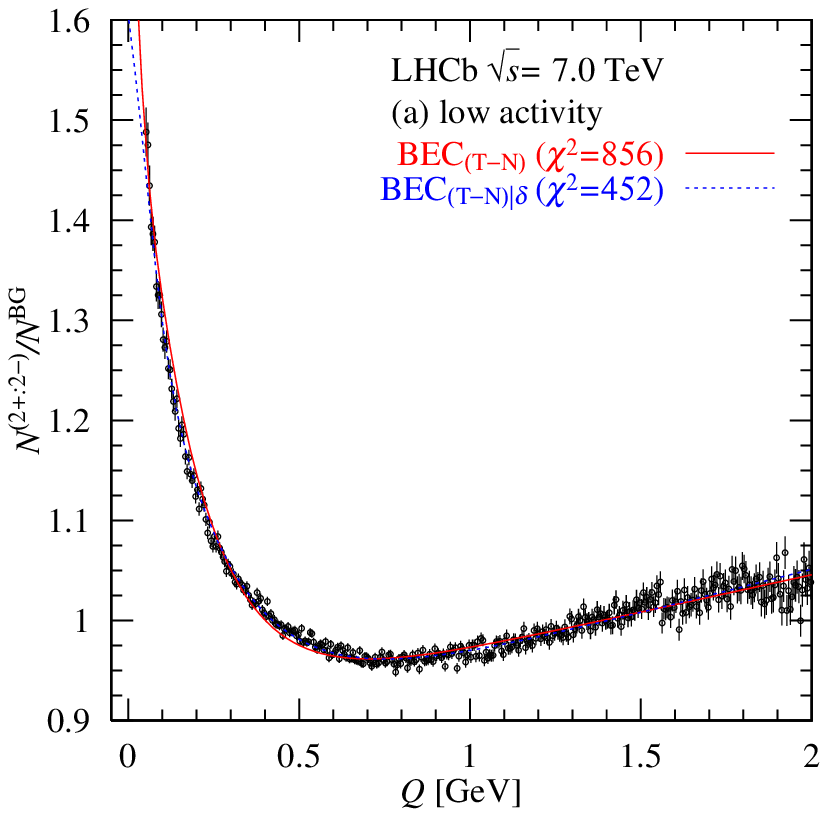}
  \includegraphics[width=0.48\columnwidth]{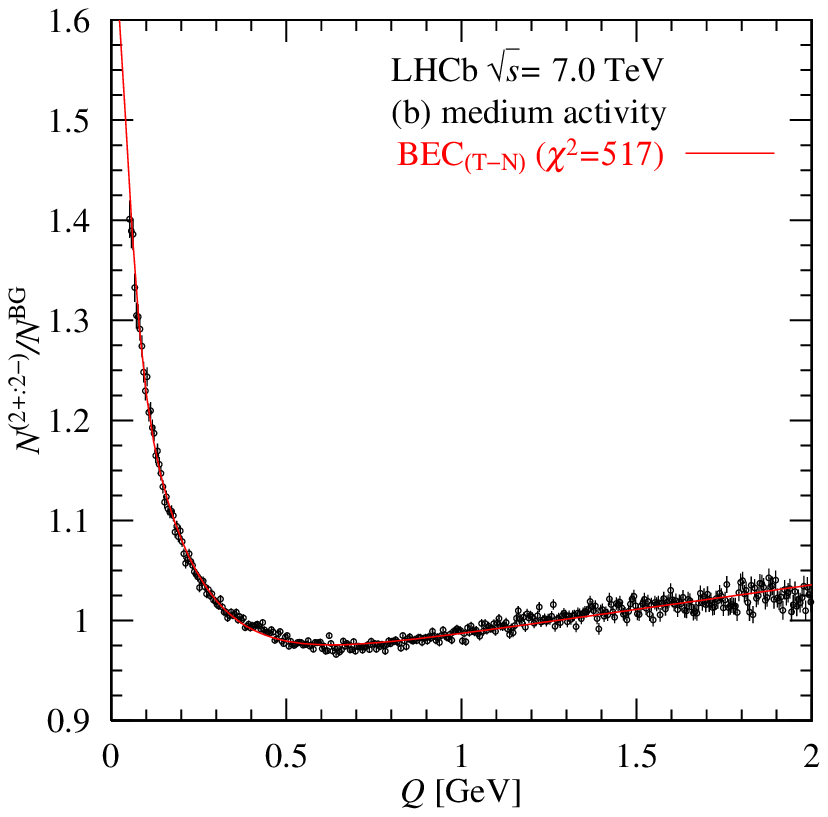}\\
  \includegraphics[width=0.48\columnwidth]{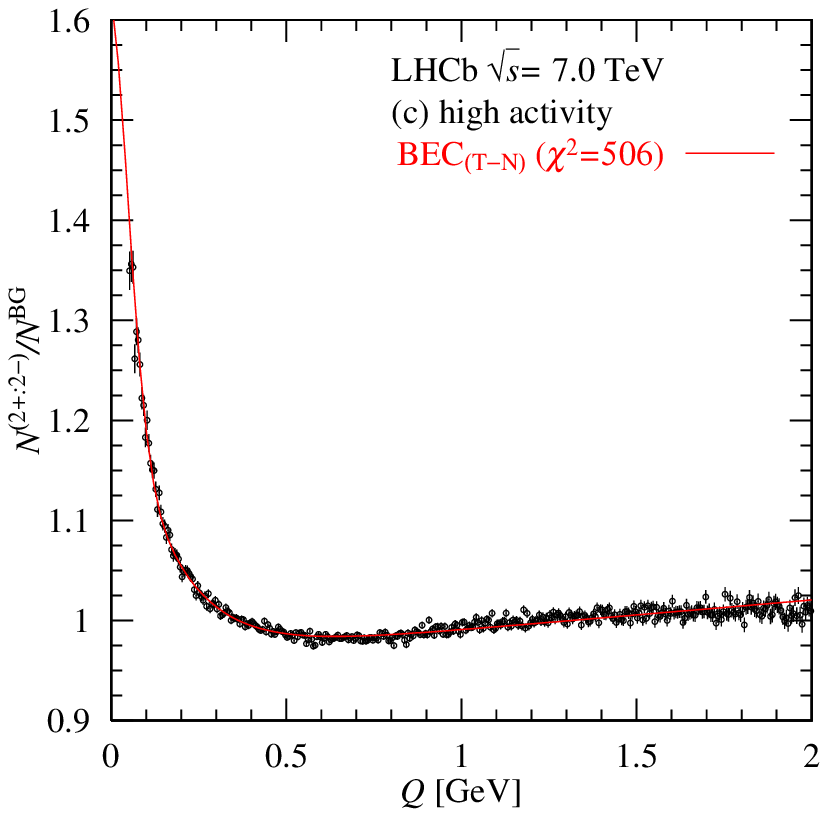}
  \caption{\label{fig4}Analysis of BEC at three activity intervals using Eq.~(\ref{eq7}) and BEC analyzed at the low activity interval using by Eq.~(\ref{eq8}).}
\end{figure}

Making use of the calculated $\lambda_i^{\rm (T\mathchar`-N)}$'s in Table~\ref{tab4} and Eq.~(\ref{eq7}), we are able to analyze the BEC by LHCb collaboration. The results of the analysis are in Fig.~\ref{fig4} and Table~\ref{tab5}.

\begin{table}[h]
\centering
\caption{\label{tab5}Results of the BEC data. $\ast)$ refers to the effective degree of coherence (see Table~\ref{tab13}). For the sake of simplicity, we assumed that $E_{\rm BE_1}=E_{\rm BE_4}$ for BEC$_{\rm (T\mathchar`-G)}$. For the three activity intervals, the long-range correlation was $\delta=(9.6\pm 0.3\sim 3.0\pm 0.1)\times 10^{-2}$ GeV$^{-1}$.} 
\vspace{1mm}
\begin{tabular}{cccccc}
\hline
Func. & $R_1$ (fm) & $\lambda_1$ & $R_2$ (fm) & $\lambda_2$ & $\!\!\chi^2$/ndf (d.e.)\\
\hline
\multicolumn{3}{l}{Low activity}\\
CF$_{\rm II}$
& $0.85\pm 0.02$ (E)
& $0.63\pm 0.01$
& $2.08\pm 0.15$ (G)
& $0.22\pm 0.02$
& $429/384$ (0.82)\\

BEC$_{\rm (T\mathchar`-N)}$
& $1.15\pm 0.01$ (E)
& 0.831 (calc.)
& $5.29\pm 0.67$ (G)
& 0.157 (calc.)
& $856/386$ (0.71)\\

BEC$_{\rm (T\mathchar`-G)}$
& $1.00\pm 0.01$ (E)
& 0.729$^{\ast)}$ (calc.)
& $3.81\pm 0.20$ (G)
& 0.116$^{\ast)}$ (calc.)
& $531/386$ (0.84)\\

\hline

\multicolumn{3}{l}{Medium activity}\\
CF$_{\rm II}$
& $1.26\pm 0.03$ (E)
& $0.50\pm 0.02$
& $2.27\pm 0.13$ (G)
& $0.19\pm 0.02$
& $495/384$ (0.79)\\

BEC$_{\rm (T\mathchar`-N)}$
& $1.35\pm 0.01$ (E)
& 0.557 (calc.)
& $3.14\pm 0.08$ (G)
& 0.270 (calc.)
& $517/386$ (0.69)\\

BEC$_{\rm (T\mathchar`-G)}$
& $1.26\pm 0.01$ (E)
& 0.501$^{\ast)}$ (calc.)
& $3.32\pm 0.07$ (G)
& 0.226$^{\ast)}$ (calc.)
& $496/386$ (0.79)\\

\hline

\multicolumn{3}{l}{High activity}\\
CF$_{\rm II}$
& $1.43\pm 0.04$ (E)
& $0.38\pm 0.02$ 
& $2.44\pm 0.12$ (G)
& $0.24\pm 0.02$
& $486/384$ (0.81)\\

BEC$_{\rm (T\mathchar`-N)}$
& $\!\!\!\!\left\{\!\!\begin{array}{l}
1.36\pm 0.02\ ({\rm E})\\
1.29\pm 0.02\ ({\rm E})
\end{array}\right.\!\!$
& $\begin{array}{l}
0.358\mbox{ (calc.)}\\
0.333\mbox{ (calc.)}
\end{array}$
& $\begin{array}{l}
2.65\pm 0.04\ ({\rm E})\\
2.58\pm 0.04\ ({\rm G})
\end{array}$
& $\begin{array}{l}
0.333\mbox{ (calc.)}\\
0.358\mbox{ (calc.)}
\end{array}$
& $\begin{array}{l}
506/386\ (0.65)\\
524/386\ (0.67)
\end{array}$\\

BEC$_{\rm (T\mathchar`-G)}$
& $1.44\pm 0.02$ (E)
& 0.392$^{\ast)}$ (calc.)
& $3.60\pm 0.07$ (G)
& 0.275$^{\ast)}$ (calc.)
& $488/386$ (0.77)\\

\hline\hline

Notes\\
\multicolumn{5}{l}{Low activity}\\

BEC$_{\rm (T\mathchar`-N)}$
& $1.90\pm 0.03$ (E)
& 0.831 (calc.)
& $0.39\pm 0.01$ (G)
& 0.157 (calc.)
& $414/386$ (0.19)\\

\hline
\end{tabular}

\end{table}

At the low activity interval, the parameters estimated by CF$_{\rm II}$ (Eq.~(\ref{eq3})) are different from those estimated by BEC$_{\rm (T\mathchar`-N)}$ (Eq.~(\ref{eq7})). At the medium activity interval, the $\lambda_1$ and $\lambda_2$ calculated by CF$_{\rm II}$ are almost the same as those calculated by T-NBD. At the high activity interval, the differences between CF$_{\rm II}$ and BEC$_{\rm (T\mathchar`-N)}$ becomes very small because the $\lambda_1$ and $\lambda_2$ calculated by T-NBD are approximately degenerated and $R_1$ and $R_2$ are almost the identical.

It should be noted that we also obtain results using double negative binomial distribution (D-NBD; Appendix~\ref{secB}). We are not able to reproduce better $\chi^2$ values using this method (see Table~\ref{tab10} in Appendix~\ref{secB}).

Moreover, several calculations made using the three-Generalized Glauber-Lachs (T-GGL) formula are presented in Appendix~\ref{secC}, and the numerical values are also shown therein. The results obtained using the T-GGL equation are included in Table~\ref{tab5}.
\\

\begin{figure}[htbp]
  \centering
  \includegraphics[width=0.46\columnwidth]{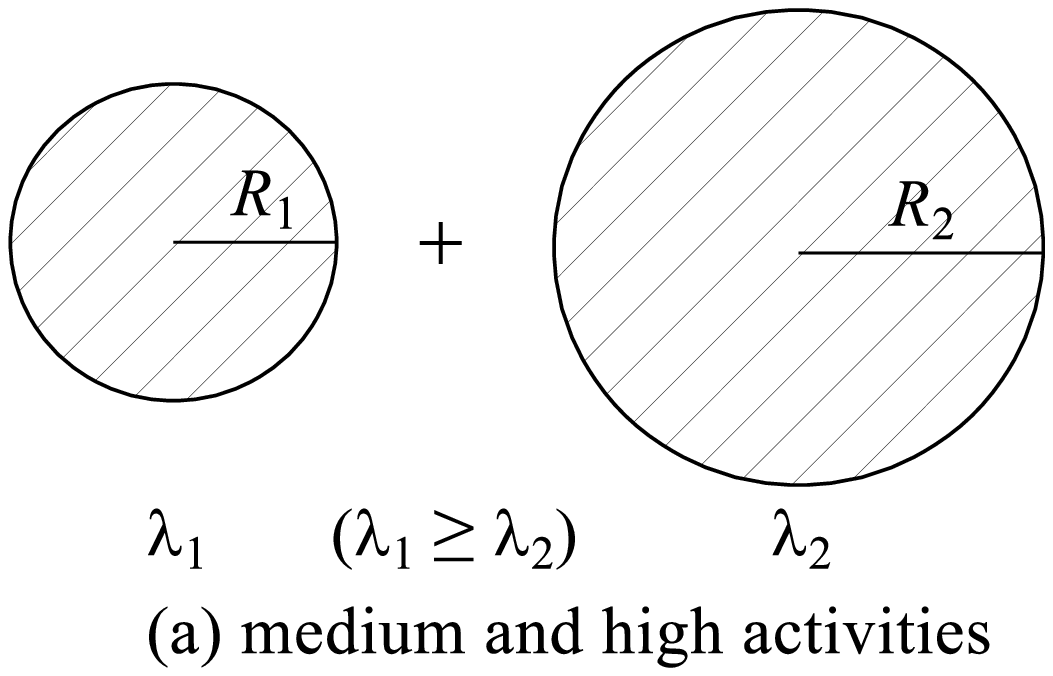}\qquad
  \includegraphics[width=0.46\columnwidth]{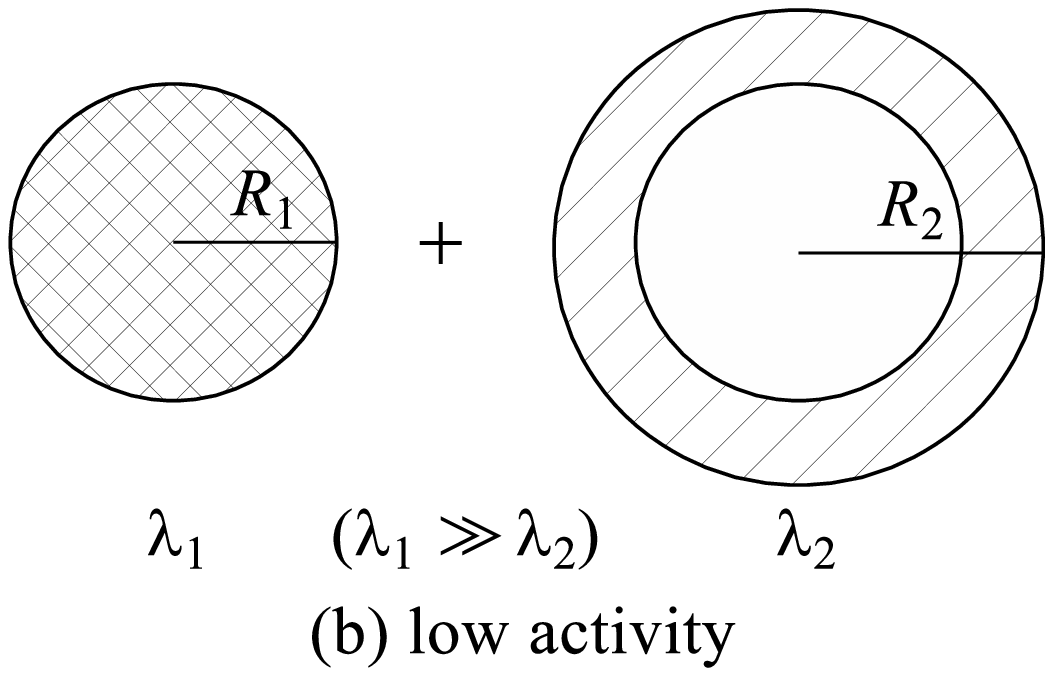}
  \caption{\label{fig5}(a) Superposition of the two production regions specified by $\lambda_{1,2}$ and $R_{1,2}$ where $\lambda_1 \ge \lambda_2$. (b) Superposition of the first region in (a) and the disk-like region with $\lambda_1 \gg \lambda_2$. The third term from the right in Eq.~(\ref{eq8}) corresponds to the disk-like region in (b).}
\end{figure}

Next, we investigated the reasons for the large discrepancies within the low activity interval shown in Table~\ref{tab5}. According to the values presented in Table~\ref{tab5}, $\lambda_1\gg \lambda_2$ is observed at the low activity interval, meaning that the production region in $\lambda_1$ differed significantly from that in $\lambda_2$. Thus, we adopt a simple subtraction method: $E_{\rm BE_2}^2 \to \delta E_{\rm BE_2}^2 =(E_{\rm BE_2}^2-E_{\rm BE_1}^2)$ for CF$_{\rm II}$ (Eq.~(\ref{eq3})):
\begin{eqnarray}
{\rm CF_{\rm II|\delta}} = 1.0 + \lambda_1 E_{\rm BE_1}^2 + \lambda_2 (E_{\rm BE_2}^2-E_{\rm BE_1}^2).
\label{eq8}
\end{eqnarray}
The results obtained by CF$_{\rm II|\delta}$ (Eq.~(\ref{eq8})) are summarized in Table~\ref{tab6}. As can be seen Table~\ref{tab6}, the coincidences among CF$_{\rm II|\delta}$, BEC$_{\rm (T\mathchar`-N|\delta)}$, and BEC$_{\rm (T\mathchar`-G|\delta)}$ are improved. Thus, differences among CF$_{\rm II}$, BEC$_{\rm (T\mathchar`-N)}$, and BEC$_{\rm (T\mathchar`-G)}$ in the low activity interval seen in Table~\ref{tab5} can be attributed to the geometrical arrangement situations shown in Fig.~\ref{fig5}.

\begin{table}[h]
\centering
\caption{\label{tab6}Results obtained using CF$_{\rm II|\delta}$ (Eq.~(\ref{eq8})) replacing $\delta E_{\rm BE_2}^2=(E_{\rm BE_2}^2-E_{\rm BE_1}^2)$ from Eq.~(\ref{eq7}), BEC$_{\rm (T\mathchar`-N|\delta)}$, and BEC$_{\rm (T\mathchar`-G|\delta)}$.}
\vspace{1mm}
\begin{tabular}{cccccc}
\hline
\multicolumn{3}{l}{Low activity}\\
Func. & $R_1$ (fm) & $\lambda_1$ & $R_2$ (fm) & $\lambda_2$ & $\!\!\chi^2$/ndf  (d.e.)\\
\hline

CF$_{\rm II|\delta}$
& $0.85\pm 0.02$ (E)
& $0.85\pm 0.02$
& $2.08\pm 0.15$ (G)
& $0.22\pm 0.02$
& $429/384$ (0.60)\\

BEC$_{\rm (T\mathchar`-N|\delta)}$
& $0.92\pm 0.01$ (E)
& 0.831 (calc.)
& $2.25\pm 0.08$ (G)
& 0.157 (calc.)
& $452/386$ (0.87)\\

BEC$_{\rm (T\mathchar`-G|\delta)}$
& $0.83\pm 0.01$ (E)
& 0.729$^{\ast)}$ (calc.)
& $2.51\pm 0.08$ (G)
& 0.116$^{\ast)}$ (calc.)
& $468/386$ (0.81)\\
\hline
\end{tabular}
\end{table}

\section{\label{sec4}Concluding remarks and discussions}

\paragraph{C1)} A second conventional formula CF$_{\rm II}$ with two degrees of coherence ($\lambda_1$ and $\lambda_2$) is proposed. This formula contains four free parameters: $\lambda_1$, $\lambda_2$, $R_1$, and $R_2$. They are determined by the MINUIT program by assigning random variables to the four parameters ($\lambda_1$, $\lambda_2$, $R_1$, and $R_2$) at the starting points. $c=0.9$ and $\delta=0.0$ in $c(1+\delta Q)$ are the initial values in our calculation.

\paragraph{C2)} We analyze data on the BEC found at 7 TeV by the LHCb collaboration using Eqs.~(\ref{fig1})--(\ref{fig4}). The results shown in Tables~\ref{tab1} and \ref{tab2} and Fig.~\ref{fig1} indicate that two degrees of coherence are necessary for analyzing BECs. The correct formula is probably associated with the physical processes of the MD with weight factors $\alpha_i$.

\paragraph{C3)} An analysis of BEC$_{\rm (T\mathchar`-N)}$ using T-NBD is presented (Eq.~(\ref{eq7})). In this case $\lambda_1^{\rm (T\mathchar`-N)}$ and $\lambda_2^{\rm (T\mathchar`-N)}$ are calculated using Eq.~(\ref{eq7}) (see Table~\ref{tab4}). In Appendix \ref{secB}, results on the BEC obtained using D-NBD are presented. D-NBD is not able to explain the 7 TeV BEC data measured by the LHCb collaboration using the numerical values shown in Table~\ref{tab10}. In the low activity interval section of Table~\ref{tab5}, $\chi^2=856$ (determined by T-NBD) is the largest value. This can be attributed to the fact that $\lambda_1=0.83$. On the contrary, in medium and high intervals, the problem mentioned above does not occur.

\begin{figure}[htbp]
  \centering
  \includegraphics[width=0.48\columnwidth]{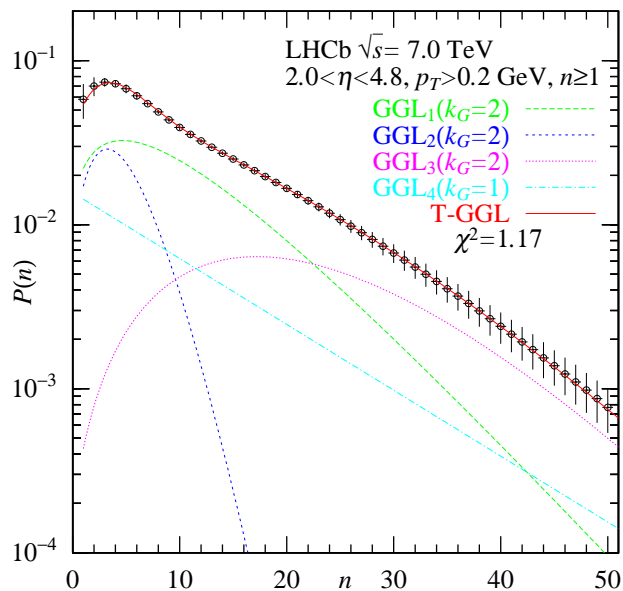}
  \caption{\label{fig6}Analysis of MD by the T-GGL including $P_{\rm GGL_4}$ with $\chi^2$/ndf$=$1.17/47. Values in Table~\ref{tab11} are used.}
\end{figure}

\paragraph{C4)} An analysis of the BEC$_{\rm (T\mathchar`-G)}$ using the T-GGL formula is presented in Appendix \ref{secC}. $\lambda_1^{\rm (T\mathchar`-G)}$ and $\lambda_2^{\rm (T\mathchar`-G)}$ are calculated by numerical values determined in the T-GGL equation in analysis of the MD (see Tables~\ref{tab5} and \ref{tab11}--\ref{tab13}). The analysis of MD by the T-GGL including $P_{\rm GGL_4}$ is presented in Fig.~\ref{fig6}.

\paragraph{C5)} The results obtained using Eq.~(\ref{eq3}) (CF$_{\rm II}$) and those obtained with Eq.~(\ref{eq7}) (T-NBD) are nearly consistent with one another. However, the results generated by BEC$_{\rm (T\mathchar`-G)}$ with calculated values based on T-GGL are nearly consistent with those generated by Eq.~(\ref{eq3}) (CF$_{\rm II}$).

\paragraph{C6)} The geometrical situation arrangements of the BECs at the three activity intervals are almost the same. An exponential function is used to describe the component with $\alpha_1$, and a Gaussian distribution is used to describe the second and third components with $\alpha_2$ and $\alpha_3$.

\paragraph{D1)} Provided that the present physical picture is correct, it can be concluded that the branching equation, including the death term, the birth term, and the immigration term, plays an important role in the collisions at the LHC. A concrete expression of this is shown in Appendix~\ref{secD}. When the continuous KNO scaling variable $z=n/\langle n\rangle$~\cite{Koba:1972ng} is introduced, the branching equation becomes the Fokker-Planck equation. From Eq.~(\ref{eq13}), the Feller stochastic process can be derived~\cite{Feller:1951aa,Biyajima:1983qu,Biyajima:1984ay}. Our analysis of Eq.~(\ref{eq15}) is presented in Fig.~\ref{fig7} in Appendix~\ref{secD}.

\paragraph{D2)} For the sake of comparison, we examine the data on BEC at 7 TeV generated by the CMS collaboration~\cite{Khachatryan:2011hi} (see Ref.~\cite{Biyajima:2019wcb}). We choose these data because of the similarity of the energy level (7 TeV) and the large number of data points (197). Our results by CF$_{\rm II}$, BEC$_{\rm (T\mathchar`-N)}$, CF$_{\rm II|\delta}$, and BEC$_{\rm (T\mathchar`-N|\delta)}$ with (d.e.) are presented in Table~\ref{tab7}. It can be seen that the $\chi^2$ value is improved compared with our previous analyses (802$\to$655).
\\

\begin{table}[h]
\centering
\caption{\label{tab7}Analysis of BEC at 7 TeV by CMS collaboration using Eqs~(\ref{eq1}), (\ref{eq3}), and (\ref{eq8}).} 
\vspace{1mm}
\begin{tabular}{cccccc}
\hline
Func. & $R_1$ (fm) & $\lambda_1$ & $R_2$ (fm) & $\lambda_2$ & $\!\!\chi^2$/ndf (d.e.)\\
\hline
CF$_{\rm I}$
& $1.89\pm 0.02$ (E)
& $0.62\pm 0.01$
& -
& -
& $738/194$ (1.0)\\

\hline

CF$_{\rm II}$
& $\!\!\!\!\left\{\!\!\begin{array}{l}
1.71\pm 0.03\ ({\rm E})\\
3.88\pm 0.18\ ({\rm E})
\end{array}\right.\!\!$
& $\begin{array}{l}
0.52\pm 0.01\\
0.84\pm 0.03
\end{array}$
& $\begin{array}{l}
3.90\pm 0.34\ ({\rm G})\\
0.71\pm 0.01\ ({\rm G})
\end{array}$
& $\begin{array}{l}
0.26\pm 0.03\\
0.12\pm 0.01
\end{array}$
& $\begin{array}{l}
612/192\ (0.7)\\
540/192\ (0.3)
\end{array}$\\

BEC$_{\rm (T\mathchar`-N)}$
& $2.07\pm 0.01$ (E)
& 0.71 (calc.)
& $6.35\pm 0.61$ (G)
& 0.12 (calc.)
& 802/194 (0.66)\\

\hline

CF$_{\rm II|\delta}$
& $1.71\pm 0.03$ (E)
& $0.79\pm 0.04$
& $3.90\pm 0.34$ (G)
& $0.26\pm 0.03$
& 612/192 (0.62)\\

BEC$_{\rm (T\mathchar`-N|\delta)}$
& $1.84\pm 0.01$ (E)
& 0.71 (calc.)
& $3.80\pm 0.21$ (G)
& 0.12 (calc.)
& 655/194 (0.73)\\

\hline

\end{tabular}
\end{table}

\noindent
{\it Acknowledgments.} T.~Mizoguchi acknowledges the funding provided by Pres. Y.~Hayashi. M.~Biyajima thanks his colleagues at the Department of Physics of Shinshu University for their kindness. 


\appendix

\section{\label{secA}Calculations with CF$_{\rm II}$ (Eq.~(\ref{eq3}))}
\paragraph{1.} We prepare four random variables for the four free parameters in CF$_{\rm II}$: $\lambda_1$, $\lambda_2$, $R_1$, and $R_2$ at the starting point of the MINUIT program.

\paragraph{2.} We perform 1,000 trials and classify them into 10 groups of 100 trials each according to the $\chi^2$ values (see Table~\ref{tab8}). 

\paragraph{3.} Finally, we observe the distribution of $\chi^2$ values in each 100-trial group to investigate the uniformity of the random variables. Table~\ref{tab8} shows that there is a clear uniformity in the data.

\begin{table}[h]
\centering
\caption{\label{tab8}Monte Carlo data of the 1,000 trials classified into 10 groups of 100.}
\vspace{1mm}
\begin{tabular}{c|cccccccccccc}
\hline
Sets   &  1 &  2 &  3 &  4 &  5 &  6 &  7 &  8 &  9 & 10 &     & (d.e.)\\
\hline
low    & 85 & 76 & 82 & 81 & 84 & 83 & 84 & 86 & 81 & 78 & 820 & (0.820)\\
Medium & 81 & 75 & 74 & 74 & 80 & 81 & 81 & 79 & 80 & 81 & 786 & (0.786)\\
High   & 82 & 73 & 83 & 78 & 85 & 83 & 83 & 81 & 75 & 83 & 806 & (0.806)\\
\hline
\end{tabular}
\end{table}

We then calculate the (d.e.) of the sets in Table~\ref{tab8} as follows: 
$$
\mbox{degree of emergence (d.e.)} = \frac{\mbox{the number of ensembles with the same }\chi^2}{\mbox{the number of trials (1\ {\rm k} =1,000)}}
$$

It is found that the first group with the largest (d.e.) corresponds to the real physical phenomena that are measured.


\section{\label{secB}Moments in three activities measured by T-NBD}
Using Eq.~(\ref{eq4}), we are able to analyze the MD at 7 TeV collected by the LHCb collaboration. Using the parameters shown in Table~\ref{tab3}, we compute several moments at each of the three activities, shown in Table~\ref{tab9}. From these values, we are able to obtain the degrees of coherence of the T-NBD results shown in Table~\ref{tab4}.

\begin{table}[h]
\centering
\caption{\label{tab9}Analysis of moments at the three activity intervals by the T-NBD shown in Table~\ref{tab3} and additional estimated parameters.} 
\vspace{1mm}
\begin{tabular}{ccccccc}
\hline
$n$ & $\langle n_1\rangle$ & $\langle n_2\rangle$ & $\langle n_3\rangle$ & 
$\langle n_1(n_1-1)\rangle$ & $\langle n_2(n_2-1)\rangle$ & $\langle n_3(n_3-1)\rangle$ \\
\hline
$[1,\,7]$   &  1.78 & 3.20 &  0.371 &   7.00 & 11.59 &   1.76 \\
$[8,\,18]$  &  4.76 & 1.40 &  4.92 &  56.2 & 12.67 &  64.5 \\
$[19,\,35]$ &  3.41 & 0.01 & 11.03 &  82.2 &  0.19 & 281 \\
$[36,\,96]$ &  0.80 & 0.00 &  5.82 &  34.6 &  0.00 & 263 \\
$[0,\,96]$  & 10.75 & 4.62 & 22.15 & 180 & 24.46 & 610 \\
\hline
\end{tabular}
\end{table}

In addition to the analysis of MD at 7 TeV using T-NBD, we perform a similar analysis using D-NBD. Using the same procedure as that used for T-NBD, we are able to obtain the final formula for the BEC. Various papers related D-NBD are referred in \cite{Fuglesang:1989st,Giovannini:1998zb,Ghosh:2012xh,Zaccolo:2015udc}.

\begin{table}[h]
\centering
\caption{\label{tab10}Analysis of the BEC by D-NBD. $(\alpha_1,\,\alpha_2)=(0.662,\,0.338)$, $(\langle n_1\rangle,\,\langle n_2\rangle)=(15.0,\,4.50)$, and $(k_1,\,k_2)=(2.18,\,3.08)$ with $\chi^2=0.12$ are used.} 
\vspace{1mm}
\begin{tabular}{c|ccccc}
\hline
Activity interval & $R_1$ (fm) & $\lambda_1^{\rm (D\mathchar`-N)}$ & $R_2$ (fm) & $\lambda_2^{\rm (D\mathchar`-N)}$ & $\chi^2$/ndf\\
\hline
Low
& $1.39\pm 0.02$ (E)
& 0.796 (calc.)
& $0.32\pm 0.01$ (G)
& 0.086 (calc.)
& 496/386\\
Medium
& $1.99\pm 0.01$ (E)
& 0.913 (calc.)
& $0.35\pm 0.03$ (G)
& 0.003 (calc.)
& 1348/386\\
High
& $2.52\pm 0.01$ (E)
& 0.916 (calc.)
& ---
& 0
& 1164/387\\
\hline
\end{tabular}
\end{table}

The results shown in Table~\ref{tab10} indicate that T-NBD does not work well for the analysis of the BEC measured by the LHCb collaboration in particular in medium and high activities.


\section{\label{secC}T-GGL formula framework}
The generalized Glauber-Lachs formula~\cite{Biyajima:1982un,Biyajima:1983qu} can be expressed as: 
\begin{eqnarray}
P_{\rm GGL}(n,\,\langle n\rangle,\,k_{\rm G},\,p) = \frac{(p\langle n\rangle/k_{\rm G})^n}{(1+p\langle n\rangle/k_{\rm G})^{n+k_{\rm G}}}
\exp\left[-\frac{\gamma p\langle n\rangle}{1+p\langle n\rangle/k_{\rm G}}\right]
L_n^{(k_{\rm G}-1)}\left(-\frac{\gamma k_{\rm G}}{1+p\langle n\rangle/k_{\rm G}}\right),
\label{eq9}
\end{eqnarray}
where $p=1/(1+\gamma)$ $(\gamma = \langle n_{\rm co}\rangle/\langle n_{\rm chao}\rangle)$ and $L_n^{(k_{\rm G}-1)}$ is the Laguerre polynomial, where $\langle n_{\rm co}\rangle$ and $\langle n_{\rm chao}\rangle$ are the average multiplicities of the coherent component and chaotic component, respectively. The T-GGL formula is expressed as
\begin{eqnarray}
P(n,\,\langle n\rangle) = \sum_{i=1}^3 \alpha_i P_{{\rm GGL}_i}(n,\,\langle n_i\rangle,\,k_{{\rm G}i},\,p_i)
\label{eq10}
\end{eqnarray}
For the charged particle distributions, we assign $k_{\rm G}^{(\pm)}=2$ to the positively and negatively charged distribution. For the identical particle distributions, we assign $k_{\rm G}^{(+)}=k_{\rm G}^{(-)}=1$ to positively and negatively charged distributions~\cite{Biyajima:1982un,Biyajima:1983qu,Shimoda:1957aa}. For the fractional parameter $k_{\rm N}=1.603$ in Table~\ref{tab3} obtained using T-NBD, we would like to adopt a T-NBD decomposition rule in the T-GGL formula, whose index $k_{\rm G}$ is specified by integers.

Using the second moment of the two frameworks, we obtain the following equation with parameters derived from NBD on the left-hand side (LHS) and GGL on the right-hand side (RHS):
\begin{eqnarray}
\frac{\langle n(n-1)\rangle}{\langle n\rangle^2} = \left(1+\frac 1{k_{\rm N}}\right) = \beta\left(1+\frac 1{k_{\rm G}^{(\pm)}}\right)+\frac{1-\beta}2\left(1+\frac 1{k_{\rm G}^{(+)}}\right)+\frac{1-\beta}2\left(1+\frac 1{k_{\rm G}^{(-)}}\right)
\label{eq11}
\end{eqnarray}
where the first term on the RHS denotes the charged particle distributions, and the second and third term terms on the RHS denote the positive and the negative distributions, respectively. When $k_{\rm N}$ is used, we obtain $\beta =0.75$. This means that the charged particle distribution given by the NBD (with $k_{\rm N}=1.6$) can be decomposed into a charged particle distribution with $k_{\rm G}^{(\pm)} =2$ and that with an identical particle distribution with $k_{\rm G}^{(+)} = k_{\rm G}^{(-)} = 1$ and the ratio $\beta=0.75$. The first component of T-NBD can be described by the following equivalent T-GGL expression with an assumption $\langle n^{(\pm)}\rangle = \langle n^{(+)}\rangle = \langle n^{(-)}\rangle$:
\begin{eqnarray}
P_{\rm NBD_1}(n,\,\langle n\rangle,\,k_{\rm N}=1.6)
 &=& \beta P_{\rm GGL}(n,\,\langle n^{(\pm)}\rangle,\,k_{\rm G}^{(\pm)}=2,\,p=1.0)\nonumber\\
 &&+ \frac{1-\beta}2 P_{\rm GGL}(n,\,\langle n^{(+)}\rangle,\,k_{\rm G}^{(+)} = 1,\,p=1.0)\nonumber\\
 &&+ \frac{1-\beta}2 P_{\rm GGL}(n,\,\langle n^{(-)}\rangle,\,k_{\rm G}^{(-)}=1,\,p=1.0)
\label{eq12}
\end{eqnarray}
The components with weight factor $(1-\beta)/2$ contain the same charged  pion ensembles: $(\pi^{+},\,\pi^{+}\pi^{+},$ $\,\pi^{+}\pi^{+}\pi^{+} + \cdots)$ and $(\pi^{-},\,\pi^{-}\pi^{-},\,\pi^{-}\pi^{-}\pi^{-} + \cdots)$. These ensembles appear to be different collections of the same pion.

Using Eqs.~(\ref{eq9})--(\ref{eq12}), we determine the values of the T-GGL parameters, which are shown in Table~\ref{tab11}. Using the values presented in Table~\ref{tab12}, we determine the degrees of coherence using T-GGL. The results are shown Table~\ref{tab13}.

\begin{table}[h]
\centering
\caption{\label{tab11}MD parameters estimated using the T-GGL equation (Eq.~(\ref{eq12})). Notice that renaming $\alpha_1^{\rm (N)}\beta = \alpha_1$ and $\alpha_1^{\rm (N)}(1-\beta) = \alpha_4$, where $\alpha_1^{\rm (N)} = 0.648$ in Table~\ref{tab3} is used.} 
\vspace{1mm}
\begin{tabular}{cccc}
\hline
$i$ & $\alpha_i$ & $\langle n_i\rangle$ & $p_i$\\
\hline
1
& $\alpha_1^{\rm (N)}\beta=0.486$
& 10.32 
& 1.000\\
2
& $\alpha_2=0.180$
& 4.42
& 0.175\\
3
& $\alpha_3=0.172$
& 22.14
& 0.260\\
4
& $\alpha_1^{\rm (N)}(1-\beta)=0.162$
& 10.32
& 1.000\\
\hline
\end{tabular}
\end{table}

\begin{table}[h]
\centering
\caption{\label{tab12}Analysis of moments at three activity intervals applying the T-GGL equation to the values in Table~\ref{tab11}, and the parameters estimated using T-GGL.} 
\vspace{1mm}
\begin{tabular}{ccccccccc}
\hline
$n$ & $\langle n_1\rangle$ & $\langle n_2\rangle$ & $\langle n_3\rangle$ & $\langle n_4\rangle$ & 
$\!\!\langle n_1(n_1-1)\rangle\!\!$ & $\!\!\langle n_2(n_2-1)\rangle\!\!$ & $\!\!\langle n_3(n_3-1)\rangle\!\!$ & $\!\!\langle n_4(n_4-1)\rangle\!\!$ \\
\hline
$[1,\,7]$   &  1.78 & 3.29 &  0.37 &  1.73 & 7.23 & 12.02 & 1.71 &   6.43 \\
$[8,\,18]$  &  5.13 & 1.31 &  4.76 &  4.04 & 60.4 & 11.48 & 62.6 &  48.0 \\
$[19,\,35]$ &  3.18 & 0.00 & 11.49 &  3.73 & 75.4 &  0.07 & 293  &  92.2 \\
$[36,\,96]$ &  0.50 & 0.00 &  5.58 &  1.80 & 20.9 &  0.00 & 244  &  85.4 \\
$[0,\,96]$  & 10.60 & 4.60 & 22.17 & 11.30 & 164  & 23.58 & 602  & 232 \\
\hline
\end{tabular}
\end{table}

\begin{table}[h]
\centering
\caption{\label{tab13}The three degrees of coherence at the three activity intervals. Effective degrees of coherence are the sum of two coefficients at the three activity intervals.} 
\vspace{1mm}
\begin{tabular}{cccc}
\hline
Activity interval & $\lambda_1^{\rm (T\mathchar`-G)}$ (calc.) & $\lambda_2^{\rm (T\mathchar`-G)}$ (calc.)\\
\hline
Low
& $0.511E_{\rm BE_1}^2 + 0.218E_{\rm BE_4}^2$
& $0.100E_{\rm BE_3}   + 0.017E_{\rm BE_3}^2$\\
Medium
& $0.303E_{\rm BE_1}^2 + 0.198E_{\rm BE_4}^2$
& $0.192E_{\rm BE_3}   + 0.034E_{\rm BE_3}^2$\\
High
& $0.123E_{\rm BE_1}^2 + 0.269E_{\rm BE_4}^2$
& $0.234E_{\rm BE_3}   + 0.041E_{\rm BE_3}^2$\\
\hline
\end{tabular}
\end{table}


\section{\label{secD}Generalized Glauber-Lachs formula}
To describe the MD at 7 TeV measured by the LHCb collaboration, we use Eq.~(\ref{eq4}) (NBD) and Eq.~(\ref{eq9}) (GGL). These are solutions to the following branching equation from laser optics~\cite{Shimoda:1957aa} and quasi-QCD~\cite{Biyajima:1983qu,Biyajima:1984aq,Biyajima:1984ay}:
\begin{eqnarray}
\diff{P(n,\,t)}{t} &=& - \lambda_0 [P(n,\,t)-P(n-1,\,t)]
+ \lambda_1 [(n+1)P(n+1,\,t)-nP(n,\,t)]
\nonumber\\
& & + \lambda_2 [(n-1)P(n-1,\,t)-nP(n,\,t)],
\label{eq13}
\end{eqnarray}
where $\lambda_0$, $\lambda_1$, and $\lambda_2$ are the immigration term (cf. ${\rm quark}\,(q)\to {\rm quark}\,(q) + {\rm gluon}\,(g)$), the death term (cf. ${\rm gluon}\,(g)\to q + \bar q$) and the  birth term (cf. $g\to g + g$), respectively. Solutions to Eq.~(\ref{eq13}) are given by Eqs.~(\ref{eq4}) and (\ref{eq9}). These make use of the initial conditions $\delta_{n,0}$ and $\langle n_{\rm co}\rangle^n e^{-\langle n_{\rm co}\rangle}/n!$ (the Poisson distribution for the coherent state), respectively.

Through hadronization (the quark-hadron duality), these become suitable candidates for describing MD with $k=\lambda_0/\lambda_2$ as follows:
$$
(q+\bar q) + (q+\bar q) + (q+\bar q) + \cdots\ \mapright{\rm hadronization}\ \pi^{(\pm)} + \pi^{(\pm)} + \pi^{(\pm)} + \cdots.
$$

\begin{table}[h]
\centering
\caption{\label{tab14}Stochastic background of NBD and GGL} 
\vspace{1mm}
\begin{tabular}{c|cc}
\hline
Equation & Initial condition & Solution\\
\hline
\lw{Branching} & $\delta_{n,0}$ & Eq.~(\ref{eq4}) (NBD)\\
equation & $\sum \dfrac{\langle n_{\rm co}\rangle^n e^{-\langle n_{\rm co}\rangle}}{n!}$ & Eq.~(\ref{eq9}) (GGL)\\
\hline
\end{tabular}
\end{table}
From Eq.~(\ref{eq13}) and the inverse Poisson transformation, we can obtain the following KNO scaling function expressed by the modified Bessel function that is a solution in the Feller process~\cite{Feller:1951aa}:
\begin{eqnarray}
\psi_{k_{\rm G}}(z,\:p) &=& \left(\frac{k_{\rm G}}{p}\right)^{k_{\rm G}} 
        \left[\frac{z}{\sqrt{z(k_{\rm G}/p)^2(1-p)}}\right]^{k_{\rm G}-1}
\nonumber\\
        && \times\exp\left[-\frac{k_{\rm G}}{p}(1-p+z)\right] 
        I_{k_{\rm G}-1}\left(2\sqrt{z(k_{\rm G}/p)^2(1-p)}\right).
\label{eq14}
\end{eqnarray}
The KNO scaling function in terms of the T-GGL formula including $\psi_{k_{\rm G}=1}(z/r_4,\,p_4)$ is expressed as
\begin{eqnarray}
\psi(z)=\sum_{i=1}^3 \frac{\alpha_i}{r_i}\psi_{k_{\rm G}=2}(z/r_i,\,p_i) + \frac{\alpha_4}{r_4}\psi_{k_{\rm G}=1}(z/r_4,\,p_4)
\label{eq15}
\end{eqnarray}
where $\sum_{i=1}^4 \alpha_i = 1.0$ and $r_i=\langle n_i\rangle/\langle n\rangle$. 

\begin{figure}[htbp]
  \centering
  \includegraphics[width=0.48\columnwidth]{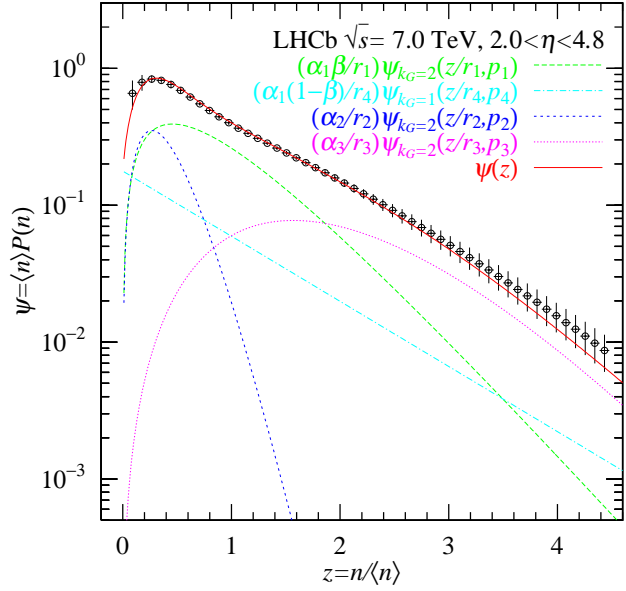}
  \caption{\label{fig7}Analysis of KNO scaling distribution by Eq.~(\ref{eq15}) with $\chi^2$/ndf$=$16.3/48.}
\end{figure}

By making use of Eq.~(\ref{eq15}) and numerical values concerning T-GGL formula, we can examine the KNO scaling distribution at 7 TeV by LHCb collaboration in Fig.~\ref{fig7}. It can be fairly well explained by Eq.~(\ref{eq15}). The coincidence between empirical data and Eq.~(\ref{eq15}) shown in Fig.~\ref{fig7} may support the usefulness of Eqs.~(\ref{eq11}) and (\ref{eq12}) mentioned in Appendix~\ref{secC}.

Finally it is worthwhile to mention a modified branching equation: No death term ($\lambda_2=0$) in Eq.~(\ref{eq13}) was studied in Ref.~\cite{Chan:1992vy}. Very recently a modified combinant analysis of that solution has been investigated in \cite{Ang:2018zjy}.


\end{document}